\documentclass[]{article}
\usepackage{stmaryrd,amssymb,amsmath,amsthm,url}
\title{Steering Fragments of Instruction Sequences}
\author{
	Jan A. Bergstra\thanks{\small Address: Science Park 904, 1098 XH, Amsterdam, 
	The Netherlands. Email: {\tt j.a.bergstra@uva.nl}. This work has been carried out with the support of the NWO Project, 
	``Thread Algebra for Strategic Interleaving''.}\\
	\newline\\
	  Section Theory of Computer Science,\\
	  Informatics Institute, Faculty of Science,\\
	  University of Amsterdam,}
	   
\date{}

\newcommand{\leftand}{~
     \mathbin{\setlength{\unitlength}{1ex}
     \begin{picture}(1.2,1.8)
     \put(-.6,0){$\wedge$}
     \put(-.53,-0.26){\circle{0.5}}
     \end{picture}
     }}
     
  \newcommand{\leftor}{~
     \mathbin{\setlength{\unitlength}{1ex}
     \begin{picture}(1.2,1.8)
     \put(-.6,0){$\vee$}
     \put(-.53,1.6){\circle{0.5}}
     \end{picture}
     }}

\newcommand{\rightand}{~
     \mathbin{\setlength{\unitlength}{1ex}
     \begin{picture}(1.2,1.8)
     \put(-.8,0){$\wedge$}
     \put(.72,-0.26){\circle{0.5}}
     \end{picture}
     }}

\newcommand{\rightor}{~
     \mathbin{\setlength{\unitlength}{1ex}
     \begin{picture}(1.2,1.8)
     \put(-.8,0){$\vee$}
     \put(.72,1.6){\circle{0.5}}
     \end{picture}
     }}

 \newcommand{\leftimp}{~~
     \mathbin{\setlength{\unitlength}{1ex}
     \begin{picture}(1.2,1.8)
     \put(-.7,0){$\rightarrow$}
     \put(-.92,0.59){\circle{0.6}}
     \end{picture}
     ~}}
     
\newcommand{\rightimp}{~
     \mathbin{\setlength{\unitlength}{1ex}
     \begin{picture}(1.2,1.8)
     \put(-.7,0){$\rightarrow$}
     \put(1.80,0.57){\circle{0.6}}
     \end{picture}
     ~}}
     
\newcommand{\leftbiimp}{~~
     \mathbin{\setlength{\unitlength}{1ex}
     \begin{picture}(1.2,1.8)
     \put(-.7,0){$\leftrightarrow$}
     \put(-.92,0.55){\circle{0.6}}
     \end{picture}
     ~}}
     
\newcommand{\rightbiimp}{~
     \mathbin{\setlength{\unitlength}{1ex}
     \begin{picture}(1.2,1.8)
     \put(-.7,0){$\leftrightarrow$}
     \put(1.80,0.55){\circle{0.6}}
     \end{picture}
     ~}}

\newcommand{\tr}{\ensuremath{T}}
\newcommand{\fa}{\ensuremath{F}}

\newcommand{\lef}{\ensuremath{\triangleleft}}
\newcommand{\rig}{\ensuremath{\triangleright}}

\theoremstyle{definition}

\newcommand{\CP}{\itname{CP}}

\newcommand{\SCL}{\itname{SCL}}

\newcommand{\itname}[1]{\ensuremath{\textit{#1}}}

\begin{document}

\maketitle

\begin{abstract}
A steering fragment of an instruction sequence consists of a sequence of steering instructions. These are decision points 
involving the check of a propositional statement in sequential logic. The question is addressed why composed propositional statements
occur in steering fragments given the fact that a straightforward transformation allows their elimination. A survey is provided of
constraints that may be implicitly assumed when composed propositional statements occur in a meaningful instruction 
sequence.
\end{abstract}

\section{Introduction}

\label{sec:1}
The occurrence of conditional statements in programs is very frequent in practice. In this paper I intend to look into the 
rationale of such occurrences  in some more detail. 
Instead of discussing computer programs in a practical program notation I will assume that programs are all presented as instruction sequences
in the notation of \cite{BergstraLoots2002}\footnote{The following notations taken from \cite{BergstraLoots2002} will be 
used without further explanation: $\#k$ for a forward jump of size $k$,   and $!$ for
a termination instruction.} or as polyadic instruction sequences in the notation of \cite{BergstraMiddelburg2008a}. Within
an instruction sequence a decision point takes the form of an instruction $+a$ or $-a$ where $a$ is a basic action. The intuition is that in an instruction sequence $X;+a;u;Y$ if $+a$ is executed the action $a$ is performed by both (optionally) changing a state and immediately thereafter
producing a boolean value (the result of $a$); then if $\tr$ is returned execution proceeds with instruction $u$ while if $\fa$ is returned that instruction is
skipped and execution proceeds with the first instruction of $Y$. In $X;-a;u;Y$ execution proceeds with $u$ after a result $\fa$ is produced (as a consequence of processing $a$) and upon the
result $\tr$ execution skips $u$ and proceeds with $Y$. 

I will call $+a$ and $-a$ steering point instructions,\footnote{In the context of this work I prefer steering point (instruction) to decision point (instruction) in order to provide for a significant distance from 
the terminology of decision making processes and methods. Further I will not use the phrase `control instructions' to maintain sufficient distance from the control code terminology of \cite{BergstraMiddelburg2009} which is strongly related to the notion
of dark programming as developed in \cite{Janlert2008}.} or steering points for short. 
Below $+a$ and $-a$ will be referred to as atomic steering points.\footnote{In \cite{BergstraLoots2002} atomic steering 
points are referred to as test instructions. This terminology is avoided in the present paper because of the risk that
it leads to confusion with the concepts of program testing and control code testing as investigated for instance 
in \cite{Bergstra2010b} and  \cite{Middelburg2010b}.} Atomic steering point are to be contrasted with non-atomic steering points,
which will constitute the central theme of the paper.
 
\subsection{Steering fragments}
A steering fragment consists of a number of steering point instructions and jump instructions. 
Jumps make use of the program counter (or rather instruction counter)
to store the decisions that were made during execution and which have been returned as boolean values by the operating environment of the instruction sequence under execution when executing steering point instructions. 
I will use the phrase `steering fragment' instead
of `decision making fragment' because the process taking care of guiding the execution of an instruction sequence is too straightforward in 
nature to justify (or require) the more general  labeling as a decision making process.\footnote{Decision making involves for instance planning, plan evaluation, formal procedures as well as their preparation, relative valuation 
of different factors and interests and perhaps organized group activity. Many papers have been written about decision making starting 
with \cite{Simon1959} where decision making itself is viewed as a task which might be profitably modeled by way of instruction sequence execution.} 

I propose to use the term steering for a simplified and strictly `programmed' form of decision making. 
In particular goals and objectives have been set out when steering begins, while
decision making may involve reflection about goals and objectives. 
In this terminology steering fragments will contain steering point instructions. 
A steering point instruction comprises a rudimentary form of decision making essentially given by the
task to evaluate a propositional statement in a setting where the evaluation of atomic propositions may have side effects. 
Besides atomic steering points non-atomic steering points are considered.
The body of a (non-atomic) steering point instruction say $+\phi$ consists of a (non-atomic) propositional statement $\phi$. For an atomic steering point
$+a$ the body is an atomic propositional statement alternatively called a propositional atom. Propositional statements are
algorithmic in the sense that a sequential order of evaluation of their parts is prescribed.

I will make use of the 
treatment of propositional statements of \cite{BergstraPonse2010} which combines the use of the notations for binary 
sequential connectives from
\cite{BergstraBethkeRodenburg1995} with the short-circuit (sequential) evaluation of the binary connectives 
often attributed to \cite{McCarthy1963} and the use of the infix conditional notation for the propositional calculus as 
proposed by C.A.R. Hoare in \cite{Hoare1985}. In particular \cite{BergstraPonse2010} is based on short-ciruit evaluation 
of the ternary conditional connective, which is made `dynamic' by allowing propositional atoms to have a side effect while 
being evaluated as in the thread algebra (polarized process algebra) of \cite{BergstraLoots2002}.
Proposition algebra is specified by the axioms  \CP\ (Conditional Propositions)  in table~\ref{CP} taken from \cite{BergstraPonse2010}.
\begin{table}
\centering
\rule[-2mm]{7cm}{.5pt}
\begin{align*}
	x \lef \tr \rig y &= x\\
	x \lef \fa \rig y &= y\\
	\tr \lef x \rig \fa  &= x\\
	x \lef (y \lef z \rig u)\rig v &= (x \lef y \rig v) \lef z \rig (x \lef u \rig v)
\end{align*}
\rule[3mm]{7cm}{.5pt}
\vspace{-5mm}
\caption{Axioms \CP\ for Proposition Algebra}
\label{CP}
\end{table}

Common binary connectives are derived from conditional composition as given by Table~\ref{DC1}. Although never used in programming practice inverse order versions of the sequential connectives can be defined in a similar fashion. Symmetric 
versions are specified in Table~\ref{DC2}. In terms of data types one may view the conditional composition as an auxiliary 
function for the definition of the sequential binary connectives. An axiomatization of the sequential binary connectives without
the help of a ternary auxiliary operator is not known to me. 

 All sequential binary connectives can be defined on the basis of 
$\{\tr,\leftand,\neg\}$, which itself is derived from \CP\ by hiding conditional composition. Using module algebra notation 
the following definition of the logic of sequential binary connectives, which I will call sort-circuit logic (\SCL)\footnote{The sequential
connectives $\leftand$ and $\leftor$ embody what is often called short-circuit evaluation of the `classical'  connectives
conjuncton and disjunction.}  is valid:

$\SCL\ = \{\tr,\leftand,\neg\}~ \Box~ (\CP\ + <\neg x = \fa \lef x \rig \tr> + <x \leftand y = y \lef x \rig \fa >)$.

In this description the conditional connective serves as an auxiliary operator for the equational specification of \SCL.\footnote{This specification of 
\SCL\ is a noteworthy example of the use of an auxiliary function in an equational specification.}
The export operator $\Box$ of module algebra hides this auxiliary operator.\footnote{As far as I know see this is in fact the most concise definition of \SCL\ available.} \SCL\ axiomatizes equivalence for propositional 
statements occurring (in the body of a steering point) in sequential programs.

\begin{table}
\centering
\rule[-2mm]{7cm}{.5pt}
\begin{align*}
	\neg x &= \fa \lef x \rig \tr  \\
	x \leftand y &= y \lef x \rig \fa\\
	x \leftor y &= \tr \lef x \rig y\\
	x \leftimp y &= (\neg x) \leftor y\\
	x \leftbiimp y &= y \lef x \rig (\neg y)
\end{align*}
\rule[3mm]{7cm}{.5pt}
\vspace{-5mm}
\caption{Negation and sequential binary connectives}
\label{DC1}
\end{table}
 
\begin{table}
\centering
\rule[-2mm]{7cm}{.5pt}
\begin{align*}
	x \rightand y &= x \lef y \rig \fa\\
	x \rightor y &= \tr \lef y \rig y\\
	x \rightimp y &= (\neg x) \rightor y\\
	x \rightbiimp y &= x \lef y \rig (\neg x)
\end{align*}
\rule[3mm]{7cm}{.5pt}
\vspace{-5mm}
\caption{Inverse order connectives}
\label{DC2}
\end{table}

Making use of arbitrary propositional statements as the body of a steering point, an instruction sequence, say, 
$X;+\phi;u;Y$ can be considered with $\phi$ some 
propositional statement. For instance, with $\phi = \neg a  \leftand  (b \rightor c)$ the following instruction sequence is obtained:

$X;+(\neg a  \leftand (b \rightor c));u;Y.$

Here $+(\neg a  \leftand (b \rightor c))$ is a steering point  (instruction) which alone constitutes an entire steering
fragment, assuming for simplicity that $u$ is not a steering instruction). Following \cite{Ponse2002} steering instructions 
containing composed propositional statements 
can always be transformed into steering fragments containing steering atoms only. This transformation produces another
instruction sequence that has the same thread as its semantics. For instance assuming that 
there are no forward  jumps from $X$ to the decision point or beyond and that there are no backward jumps from $u;Y$ to 
the decision point or before the following transformation works:

$X;+a;\#5;+b;\#2;+c;u;Y.$

\noindent An instruction sequence with a longer steering fragment (consisting of 4 instructions at least)  is:

$X;+((\neg a \leftor \neg c) \leftand (b \leftor (c \leftand d)));\#5;-((b \rightand \neg c) \rightand d);\#2;u;v;w;Y$

\noindent Again it is easy to transform this instruction sequence into an equivalent one only containing atomic steering points
using the techniques outlined in \cite{Ponse2002}.

\subsection{Questions about non-atomic steering points}
By itself the existence of steering fragments inside instruction sequences is plausible and unproblematic. Nevertheless a number
of  questions can be posed about such fragments which justify further investigation. The series of question listed below is not
meant as  a detailed plan of action for this paper. Rather the objective for writing this listing is to indicate an collection of issues
that constitute a context for further reflection. Readers may decide for themselves to what extent the mentioned issues have 
been properly, and perhaps in some cases conclusively, addressed in the sequel of the paper.
\begin{enumerate}
\item The notion of a steering point instruction underlies that of a steering fragment. Non-steering fragments might be called 
working fragments. Working fragments do not contain steering points but may contain jump instructions. 
Is this distinction sufficiently clear or would it be 
preferable to allow some occurrences of steering points as parts of working fragments as well.
\item I will assume that most programs can be transformed into (polyadic) instruction sequences ins a rather 
straightforward fashion following the projection semantics of \cite{BergstraLoots2002}. It is reasonable to translate the decision points of conditional statements into steering points containing corresponding propositional statements. Having these translations at hand it is possible to talk about the occurrence of steering fragments and steering points in a program (before translation) 
by referring to such fragments and instructions in its translation instead. Now assuming that these translations are performed 
on a large scale it may be asked how frequently non-atomic steering points that is steering points with composed propositional statements as their body, occur in practical imperative programming.

I will not provide an answer to this question, instead I will assume that non-atomic steering points 
occur quite frequently in programming practice.
\item Given the fact that all instruction sequences can be translated into semantically equivalent instruction sequences
in which steering points are atomic, why are non-atomic steering points frequently used in practice.
\item Non-atomic steering points contain propositional atoms as components, these are ordinary basic 
actions of which the boolean reply is being used: are there any plausible constraints that should be or might be imposed 
on the basic actions that occur as steering atoms? For instance restrictions on the impact of their side-effects.

\item  Non-atomic steering points contain propositional statements in a proposition algebra based on reactive valuations as described in \cite{BergstraPonse2010}. Different semantic models for this proposition algebra exist and each model codifies
a potential framework of constraints on the side effects and the results of propositional atoms. By transforming propositional statements to a normal form characteristic for a particular model instruction sequences may be made more robust against 
model changes. 

This robustness works as follows: an instruction sequence, and in particular its non-atomic  steering points, are robust against a modification of the semantics of propositional statements if that change does not, or only in a minor way suggest a modification 
of the instruction sequence (or the non-atomic steering points it contains) in order to comply with its designer's objectives in
the modified circumstances. How to cary out a satisfactory measurement of robustness is yet another question of course.

\item Assuming an execution architecture with target services and auxiliary (or local, or called para-target) services a (polyadic) instruction sequence 
for that architecture can be considered useful for a certain purpose. The corresponding assertion of usability is by itself an assertion about the instruction sequence, the architecture, its operating context and the application at the same time. If the 
context changes the usability assertion may cease to be valid and a modification (often called maintenance) of the instruction sequence may be in order to restore its validity. Are there convincing cases where maintenance of this nature is limited to dealing with the consequences of modified semantics for propositional statements (by allowing less constrained reactive 
valuations for instance).
\item In \cite{BergstraPonse2005, BergstraBethkePonse2007} the viewpoint is put forward that the well-known argument 
about the undecidability of 
viral presence in programs as claimed by \cite{Cohen84} is compromised by a lacking account of the side-effects of 
evaluation of a hypothetical steering atom. The proof of its nonexistence in \cite{Cohen84} seems to overlook the possibility
that the very property (virality) about which this steering atom is supposed to be informative is sensitive to the side effect
of the steering atom which solely consists of it having been performed together with the corresponding increment of the program counter. It seems reasonable to distinguish internal dynamics of a steering point from its external dynamics. The internal
dynamics relates to the side effect which the execution of a steering point instruction has on the program counter, while the
external dynamics concerns side effects on the services provided by an execution architecture. Proposition algebra is geared towards a description of external dynamics of non-atomic steering points. The issue with the undecidability of virus presence
concerns internal dynamics, however. It is unclear to what extent internal dynamics of non-atomic steering point instructions can be analyzed by means of sequential propositions with dynamic semantics in a proposition algebra setting.
\end{enumerate}

\subsection{Motivation and justification of the work}
This work has a dual motivation/justification. On the one hand previous work on proposition algebra (\cite{BergstraPonse2010})
calls for further
reflection concerning the role of non-atomic steering points in imperative programming. Indeed the main justification 
of sequential propositional logic (also called short-circuit logic) that underlies proposition algebra
is based on McCarthy's observation in \cite{McCarthy1963}  that
short-circuit evaluation evidently and unambiguously is the natural interpretation of well-known binary logical connectives, in
particular conjunction and disjunction, in the context of imperative programming. I am inclined to add to this that these logical 
connectives
primarily feature in imperative programming, which currently seems to be far more widespread than formal logical and mathematical reasoning which is conventionally considered to constitute the proper niche of propositional calculus.

On the other hand the option to use non-atomic steering points seems to contribute significantly to imperative
programs (here instruction sequences) as a means of algorithmic expression. If this holds true, a proper understanding of the
relation between proposition algebra and instruction sequence semantics will contribute to a further understanding of the
concept of an imperative program, even if only viewed as a means of algorithmic expression.

\section{Semantic aspects of non-atomic steering points}
Let $+\phi$ be a non-atomic steering point. Here $\phi$ is a non-atomic propositional statement. The evaluation of propositional
atoms inside $\phi$ takes place in a sequential and predetermined order where side-effects that influence subsequent 
evaluations are not excluded. Different semantic models for proposition algebra as discussed in \cite{BergstraPonse2010} correspond with different degrees of freedom concerning side-effects. 

\subsection{Reactive valuation class semantics}
A reactive valuation specifies how a succession of evaluations of propositional atoms takes place, where each atom may
cause a state change and generates a boolean reply (reaction) depending on the state in which it has been performed.
Two propositional statements can be called equivalent if their effect on all reactive valuations in some class $C_{rv}$ 
coincides. Given class $C_{rv}$ there is always a smallest congruence on propositional statements contained in this equivalence 
relation and that congruence determines a model of proposition algebra directly derived from the class $C_{rv}$.

\subsubsection{Reactive valuation classes: a survey}
Here is a non-exhaustive survey of some semantics models for proposition algebra obtained from particular classes 
of reactive valuations, indeed many more models can be designed:

\begin{description}
\item{\em Static valuation semantics}. In static valuation semantics no side-effects are permitted. Static semantics corresponds to 
ordinary propositional calculus.
\item {\em Memorizing valuation semantics}. In memorizing valuation semantics once a steering atom has been executed until
a work atom is performed subsequent
executions do not change the state and generate the same boolean outcome. 
\item {\em Weak positively memorizing valuation semantics}. In weak positively memorizing semantics after a steering atom has been evaluated to a positive outcome (i.e. $\tr$), as long as all subsequent steering atoms have positive results as well, a further
occurrence of the same steering atom must lead to a positive outcome.
\item {\em Weak negatively memorizing valuation semantics}. In weak negatively memorizing semantics after a steering atom has been evaluated to a negative outcome (i.e. $\fa$), as long as all subsequent steering atoms have negative results as well, a further
occurrence of the same steering atom must lead to a negative outcome.
\item {\em Contractive valuation semantics}. In contractive valuation semantics it is assumed that in the case of successive evaluation of the same steering atom the second evaluation does not change the state and generates the same reply as the 
first execution. In this case repeated executions of the same steering atom can be contracted to a single execution.
\item {\em Repetition proof valuation semantics}. In repetition proof valuation semantics it is assumed that in successive
executions (that is repeated evaluation) of the same steering atom the second execution generates the same boolean 
reply as the first one. In \cite{BergstraPonse2010} it is shown that in repetition proof valuation semantics (and for that reason also
also in free valuation semantics) the two-place 
connectives together have less expressive power than the three place conditional connective.
\item {\em Free valuation semantics}. In free semantics no restrictions on side-effects are assumed. 
In \cite{BergstraPonse2010} it is shown that in free valuation semantics the two-place 
connectives together have less expressive power than the three place conditional connective. The proof for free valuation semantics follows from but is far simpler than the proof in the case of repetition proof valuation semantics.
\end{description}

Different models of proposition algebra combine different mixtures of the following forms of
restrictions on the interference of the evaluation of atomic propositions. 
\begin{itemize}
\item Commutation rules that allow the order evaluation of pairs of atoms to be interchanged while preserving side-efects and boolean replies.
\item Reply memorization rules which indicate that under certain restrictions if a propositional atom is evaluated twice during
the evaluation of a propositional statement the second evaluation must produce the same reply as the first one.
\item Contraction rules which impose that under certain restrictions if a propositional atom is evaluated twice during
the evaluation of a propositional statement the second evaluation must produce the same reply as the first one and in addition 
will have no side effect (which implies that the evaluation can be avoided).
\end{itemize}

Deviations from ordinary propositional calculus appear if the evaluation of propositional atoms fails to commute or if successive
evaluations (with perhaps other evaluations in between) lead to different replies.
A propositional statement is called non-repetitive if for no reactive valuation its evaluation leads to the repeated execution 
of any propositional atom with or without the intermediate evaluation of other atomic propositions. 
For example $a \leftand (b \leftor a)$ is repetitive and $a \leftand (b \leftor \tr)$ is non-repetitive.

Concerning non-repetitive propositional statements the following information is available:
\begin{description}
\item{\em Terms with disjoint atoms are non-repetitive}. In particular the following propositional statements are
non-repetitive: $x \lef y \rig z$, $x \leftand y$, $x \leftor y$, $x \rightand y$, $x \rightor y$ and if $t$ is non-repetitive the 
so is $\neg t$.
\item {\em Conditional composition is an essential primitive}. One of the obvious ways to express the 
conditional connective in terms of 
two-place connectives is as follows $x \lef y \rig z = (x \leftand y) \leftor (\neg x \leftand z)$. However, the second expression is repetitive. Indeed from \cite{BergstraPonse2010} one may extract as a simple corollary of the results presented that a 
non-repetitive expression for $x \lef y \rig z$ does not exist. For this reason the conditional connective plays a significant 
role in steering point instructions.
\item {\em Memorizing valuations semantics allows repetititive expressions}. The definition $x \lef y \rig z = (x \leftand y) \leftor (\neg x \leftand z)$ is valid in static valuation semantics and in 
memorizing valuation semantics whereas it fails in the other semantic models mentioned above. 
In both cases repeated execution of a propositional atom has no side-effect and returns 
an equal value so that the non-repetetiviness of the right-hand side expression is immaterial.
\item {\em Eliminating repetitions cannot be done efficiently}. Using the conditional connective and the 
constants $\tr$ and $\fa$ all propositions can be written in a non-repeating form which is equivalent in 
memorizing valuation semantics.
This form is often referred to as a BDD. A non-repeating expression involving $x \lef y \rig z$, $\tr$, and, $\fa$ 
only (provided it has been normalized in the sense that boolean constants are not at the root of conditionals) 
is satisfiable if and only if it contains a constant $\tr$. Therefore unless P = NP, a transformation bringing 
all propositional statements in non-repetitive form cannot be performed in polynomial time.
\end{description}

It can be concluded that repetitive propositional statements can be turned into non-repetitive equivalent ones only in sufficiently
abstract models of proposition algebra and certainly not in the most general case, free valuation semantics. 
Even if transformation of repetitive propositional statements to non-repetitive form can be done effectively the task
cannot be done efficiently (unless P = NP).

\subsubsection{Reply stable evaluation of propositional statements}
For non-constant propositional statements the evaluation process can develop in different ways. If each atom is evaluated
with the same boolean as a result in each of its individual executions the overall evaluation of the propositional statement is
called {\em reply stable}. If the evaluation of a propositional statement consisting the body of a steering point instruction is 
reply stable that will be called a {\em reply stable evaluation of a steering point instruction}.

The fact that an evaluation is not reply stable can be observed during the evaluation to the extent that for instance an 
exception can be raised as soon as it happens. Once during the evaluation of, say, $\phi$ 
a propositional atom $a$ is evaluated returning a boolean value that differs from a previous evaluation if $a$ the fact that
the evaluation is not reply stable has been established and will remain valid during all further steps of the evaluation of $\phi$.
This differs notably from side-effects of the execution (evaluation) of atomic propositions. Such side-effects can be observed 
only indirectly, and indeed only by considering a number of evaluations in different states. Reply stability of steering 
point evaluations  is what the execution architecture of an instruction sequence can monitor during its activity, while the
absence of side-effects is a matter of abstract modeling of the sate space. If the abstract model is unknown it is difficult for
the execution architecture to make a `guess' and on that basis to produce assessments about the presence or 
absence of side-effects. 

Clearly non-reply stability itself witnesses the presence of side-effects provided no other external forces are at work.
In a real time situation the change of a reply need not be caused with the evaluation of any previous atomic proposition,
instead it may witness the fact that evaluation involves real tome reading of sensorial data in a dynamic setting. 
But it is difficult to determine which actions precisely have had side-effects, and which side-effects can be considered the 
cause of an observed fluctuation of replies.

Using the notion of a stable evaluation advantages and disadvantages of the use of repetitive and non-repetitive 
propositional statements can be surveyed. 
\begin{itemize}
\item Every evaluation of a non-repetitive propositional statement is always reply stable.
\item If a non-stable evaluation of a repetitive propositional statement takes place
 the `design logic' of the instruction sequence may be compromised in the sense that the meaning of
the propositional statement that constitutes the body of a steering instruction cannot be properly understood 
by means of its common understanding within static semantics.
\item If the evaluation of a repetitive propositional statement features a changing value of some atom it must be decided
whether or not it is meaningful to re-evaluate the entire statement from the beginning once more under the assumption 
the the renewed evaluation has a better chance of being completed in a reply stable fashion.
\item If the exception of non-reply stability is handled by renewed evaluation of an entire propositional statement the
question arises how often that way of handling the exception must be repeated if the exception itself reappears one 
or more times.
\item If all steering point instructions of a polyadic instruction sequence 
are written in a non-repetitive form a change of the semantics of propositional 
statements that underlies the execution architecture will not by itself be an incentive to redesign the instruction sequence.
No new (or old) anomalies can be detected during evaluation. 
\item Instruction sequences with non-repetitive steering points are robust against modification of their semantics. For that
reason the use of  non-repetitive steering points seems to be preferable from an instruction sequence design point of view.
In general the conditional connective `$x \lef y \rig z$' will be needed to write steering points in a non-repetitive form.
\end{itemize}

These considerations suggest an empirical question: how frequent is the existence of repetitive steering points in practice.
Not very frequent so it seems. But we have not made any systematic investigation of this matter. Nevertheless, there seems to
be ground for the following hypothesis:

\begin{quote} In the practice of instruction sequence design non-atomic steering points of moderate complexity are 
preferred when leading to shorter instruction sequences but the use of repetitive propositional statements in steering 
point instructions is systematically avoided.
\end{quote}

\section{Pragmatic aspects of non-atomic steering points}
In view of the fact that when designing an instruction sequence it is an easy task to do away with all non-atomic steering 
point instructions by means of straightforward transformations, and of the conclusion of the previous section that
non-repetitive steering points are to be preferred (with atomic steering points being non-repetitive by definition) 
the presence in practice of non-atomic steering points requires further explanation. 

\subsection{Advantages of the use of non-atomic steering point instructions}
Here are some advantages of the use of non-atomic steering points. 
\begin{itemize}
\item Instruction sequences can be made shorter if non-atomic  steering instructions are used.
\item Steering fragments have a more clear meaning because each position within a steering fragment stands 
for a larger propositional statement that can be obtained from the individual steering instruction bodies by means of
conjunction, disjunction and negation. The complexity of this larger propositional statement grows exponentially in the 
length of the steering fragment, because of the exponential growth of the number of computation paths through the steering fragment. With shorter steering fragments that may be within reach when using non-atomic steering points 
 this explosion bites less.
\item If  semantic analysis of an instruction sequence is performed using Floyd type inductive assertions, or, provided an 
instruction sequence was written using structured programming primitives (see \cite{BergstraLoots2002}), by means of some 
Hoare logic the number of intermediate assertions grows linearly with the number of instructions. Therefore reducing the number
of instructions is useful in principle.
\item Larger propositional statements in steering instructions may convey meaning which is far more easily
understandable  for a human software engineer than a multitude of atomic propositions wrapped in an instruction sequence
made up from atomic steering instructions and jumps. 
In particular top-down design methods working from a high level specification may give
rise to the use of non-atomic steering instructions.
\item Related to this matter the presence of non-atomic steering points may simplify reverse engineering of an instruction
sequence  into  specifications and it may be helpful for the effectiveness of  optimizing compilers. After all removing non-atomic
steering instructions is a trivial operation for a compiler and the added value of having done that already is minimal.
\end{itemize}

In principle the use of non-atomic steering points in instruction sequence construction can be investigated empirically by
inspecting software libraries. My objective is different, however, because I intend to perform a qualitative investigation of the arguments that come into play.

\subsection{Minimizing the size of an instruction sequence}
Given an instruction sequence $X$ and assuming memorizing valuation semantics for propositional statements the question 
may be posed to find a shortest instruction sequence with an equivalent semantics. I assume that thread extraction as used in
\cite{BergstraLoots2002,BergstraMiddelburg2007,BergstraPonse2007} is used to determine instruction sequence semantics.

In view of the preference for  a restriction to the use of non-repetitive steering points as stated above it can be
taken as an additional constraint that all steering points after optimization are non-repetitive.
Measuring the length of an instruction
sequence requires a software metric tailor made  specifically for instruction sequences. A survey of classical software
metrics is found in \cite{Kafura1985}. Using LOC (lines of code, see for instance \cite{Jones1994}) as a metric works for the present purpose  provided at most a single instruction is placed on 
each line and under the assumption that non-atomic steering instructions are written on consecutive  lines in a systematic fashion.
Even simpler, given an appropriate ASCII syntax for instructions and instruction sequences the number
of characters is a reasonable measure.\footnote{This metric is less abstract than any of the metrics surveyed in \cite{Cook1982}.
Perhaps it does not even deserve the name for that reason. In fact developing a theory of instruction sequence metrics seems 
still to be a challenge in spite of the formidable amount of work on software metrics in existence already.} 
Then some encoding for the logical connectives must be assumed. I assume that
such is done to the effect that all basic actions and all connectives as well as `$;$',  `$!$', `$($', and `$)$' have unit size.
Then the following can be concluded:
\begin{enumerate}
\item Finding a shortest equivalent instruction sequence which features non-repetetitive steering instructions only, creates
a combinatorial explosion. Indeed consider the instruction sequence $+\phi;\#3;a;!;b;!$. It corresponds with the thread
$a \lef \phi \rig b$. It can only be brought into a form with non-repetitive steering instructions by writing the propositional
statement $\phi$ in such a form. This is technically trivial but if it can be done in polynomial time P = NP.

Using an encoding of an NP complete problem in a single propositional statement as is done in \cite{BergstraBethke2010}
one obtains an argument that this combinatorial explosion not only involves time but the size of the resulting
propositional statement as well.

\item The shortest instruction sequence equivalent to say $X= +(a \leftand b);c;!$ contains a non-atomic steering point. Indeed
$-a;!;-b;c;!$ has a size of 11 while $X$ has size 10.

\item Minimizing the number of instructions is a difficult task. Indeed if $\phi$ is not satisfiable $+\phi;a;b;!$ is equivalent to
$-\phi;b;!$ but finding this out is NP hard.

\item The computational complexity of minimizing the size of an instruction sequence modulo equivalence is at most 
exponential time in its size.

\item For the above reasons in general reasonably short instruction sequences are likely to involve non-atomic steering 
points. If a software engineer insists on using non-repetitive steering points only, the number of instruction used is likely to 
increase above the theoretical minimum unless unreasonably sized steering points are accepted.
\end{enumerate}

\section{Constraints for steering point atoms}
Propositional atoms that may occur in an atomic or non-atomic steering point instruction will be called steering point atoms.
Let $A_{sp}^{c}$ consist of the propositional atoms (basic actions) which plausibly occur as constituents of atomic and 
non-atomic
steering points in instruction sequencing context $c$. Thus $A_{sp}^{c}$  denotes the steering point atoms in a certain 
context $c$. $A_{sp}^{c}$ is a subset of the collection of basic actions $A^{c}$ available for (polyadic)  instruction sequence construction in a particular context $c$. The question to be confronted here is what may be meant by plausibility in this context. 
Here are some rules concerning that matter, presented in decreasing order of priority. In the rule \ref{Rule} a more detailed 
analysis is required and individual occurrences of propositional atoms are classified as steering atom occurrences or or 
work atom occurrences.
\begin{enumerate}
\item {\bf Non-trivial boolean results.} If $a$ always returns the same boolean value (either $\tr$ or $\fa$) then it is 
implausible for $a$ to be 
included in $A_{sp}^{c}$. The reason for this rule is that in this case execution of $a$ cannot directly influence the steering of 
program execution. The only rationale of its execution lies in anticipated side-effects on the state of the particular service 
responsible for executing the action. In other words executing $a$ is better classified as work, while steering point atoms 
must, in principle, be capable of making a distinction between different states.\footnote{Some Java compilers 
seem to enforce this restriction by disallowing steering instructions of the form $+\tr$ and $+\fa$.}

\item {\bf Trivial side-effects.} Complementary to the case where the boolean reply is `trivial', if a basic action never has a non-trivial side effect and it may return different boolean values its 
inclusion in $A_{sp}^{c}$ is very plausible. These are pure observations meant not to interfere with the state in any way. 

This rule
has a lower priority than the previous rule so that a basic action (often denoted as {\tt skip} or {\tt nop}) which combines trivial 
side-effects with a trivial result is classified outside the steering atoms as a work atom.

\item {\bf Marginal side-effects.} For a basic action to qualify as a steering atom on other grounds than on the basis 
of the two preceding rules,  it is preferably required (that is, it counts as
good programming style) that its execution is not required for obtaining the correct output of instruction sequence execution.

That is, if by magic the boolean value which its execution produces would be available at the right moment during a run, available
for steering the course of further activity, a valid output would be produced if the action is left unperformed and this boolean
reply (made available be magic) would be used to steer the computational process instead. Application of the rule of
`marginal side-effects' is far more arbitrary than application of the previous rules because it depends on the 
functionality to be implemented as well as on appropriate abstraction levels for its description. This is further detailed in 
two remarks:
\begin{itemize}
\item Semantic formalization of this rule involves the introduction of an equivalence relation on states where the required output
of instruction sequence execution is specified modulo this equivalence only and where the side-effect of steering atoms
will always preserve equivalence of states. Thus while the boolean result of a steering action may not be the same for 
equivalent states the corresponding side-effects must respect equivalence of states.
\item In contrast with the previous rules this rule is specific for a particular polyadic instruction sequence. Given an 
instruction sequence, and requirements for its functionality an attempt can be made to introduce an equivalence relation on states which both satisfies the constraint that outputs of computations of the polyadic instruction sequence are only 
specified (given said requirements) modulo that equivalence and the constraint that it helps to classify some 
basic actions as having marginal 
side-effects thus providing a justification for their inclusion in $A_{sp}^{c}$. In this setting the requirements of the
functionality to be provided by running the instruction sequence is considered to constitute a part of the context $c$.
\end{itemize}

\item {\bf Marginal side-effects by default.} For some basic actions the side-effects may be classified as marginal in the sense
of the previous rule in normal cases whereas in some exceptional cases a non-marginal side-effect can be observed. In
such cases classification of a basic action as a steering atom is justified. Clearly this justification requires the provision of a 
distinction between normal and exceptional circumstances during an execution. This aspect introduces additional 
arbitrariness in the classification mechanism. Nevertheless in some circumstances (given by functional requirements, 
execution architecture and the polyadic instruction sequence at hand) the distinction between the 
normal case and the exceptional case may 
be straightforward and convincing.

\item {\bf Marginal side-effect generating occurrences.} \label{Rule} If the side-effects of carrying out a basic action are 
not always marginal, an attempt may
be made to distinguish two cases: executions with marginal side-effects (to be accepted as part of the evaluation of steering 
atoms) and execution that cause non-marginal side-effects instead. Here the distinction of a subclass $A_{sp}^{c}$ of the
basic action collection $A^{c}$ is unconvincing and the classification needs to be made at the level of occurrences in 
instruction sequences instead. The plausibility of propositional atom occurrences in steering points now depends on the ability
to distinguish occurrences where their side-effects are marginal, while accepting that in other occurrences 
side-effects may not be marginal.

\item {\bf Marginal side-effect generating occurrences by default.} The fact that a basic action occurrence within a steering 
point instruction generates marginal side-effects may hold true in some normal case only. Once a convincing demarcation 
between normal and exceptional has been proposed this state of affairs can be put forward as a justification of using
the action within a steering point instruction.

\item {\bf Detectible side effects.} A side effect is detectible by some steering atom if the fact that the side-effect has 
occurred can be 
measured by means of a steering instruction containing that atom. If side-effects of steering point atoms 
are detectible several different 
constraints can be formulated. 
\begin{itemize}
\item Side effects of steering point atoms cannot be detected by other steering point atoms in the same instruction sequence.
\item Side effects of steering point atoms cannot be detected by other steering point atoms which occur in the same
propositional statement.
\item Side effects of steering point atoms can be detected by other steering point atoms that occur in instruction sequences
that run concurrently in a multithread setting governed by some strategic interleaving (see \cite{BergstraMiddelburg2007}).
\item Instruction sequences are classified as `applications' or `system utilities'. Now steering point atoms in applications can 
only be detected by steering point atoms occurring in system utilities (which are supposed to execute concurrently
in strategic interleaving).
\end{itemize}

\end{enumerate}
The rules just mentioned may explain to a great extent which occurrences of steering point atoms are plausible. Empirical 
survey research on a large body of practical programs may be needed in order to find out about the validity of these rules
in relation to past and current imperative programming practice.

\section{Non-atomic steering points in real time systems}
Leaving aside side effects of steering atoms the only way in which non-reply stable evaluations of steering points
can take place during the execution of an instruction sequence results from real-time phenomena. This is the most
plausible cause of non-reply stability for steering point evaluations that I can imagine.

If a steering atom
provides information about a measurement made in the external world modification of the reply on the same atom
during the evaluation of a simple propositional statement is plausible provided the evaluation is sufficiently slow
in relation to the external dynamics. An example can be found in decision making on whether or not to pass a traffic 
light depending on its color (see for instance \cite{MahalelZaidelKlein1985}).

In a real time context sensor data or other measurements such as stock prices or interest rates take values in  a meadow
(see for instance \cite{BethkeRodenburg2009}). Functions taking one or more element of a meadow into a boolean produce
the meaning of propositional atoms. For instance $a \equiv {\tt height > 3000 ~meter}$ can be used as a propositional atom.
It is fairly easy to design an example where
the same meadow valued external input is used more than once (though in different atomic propositions) within the same propositional statement which itself is
used in a meaningful steering point  instruction. However, I have not yet found a convincing example where the same 
atomic proposition occurs in such a way that the propositional statement involved becomes repetitive. Such an example 
might provide in principle a most convincing illustration of the use of repetitive propositional statements in steering point instructions.
A further challenge is to find a convincing example where the transformation of steering point  instructions to their 
non-repetitive form is unfeasible from a complexity point of view. It is far from clear that such an example exists. It is
perhaps implausible that such an example can be derived from today's programming practice.

Deriving final conclusions from the above is hardly possible, 
but some hypotheses can be put forward given the qualitative work of this paper:
\begin{enumerate}
\item Reasonably small sized instruction sequences (with respect to their functionality) 
are likely to contain non-atomic steering point instructions.
\item Compilers should not be expected to transform instruction sequences with non-atomic steering points into instruction sequences featuring only atomic steering points without leading to an increase of the number of instructions.
\item \SCL\ provides a specification of the meaning preserving transformations of propositional statements inside steering points.
\item Current programming practice allows to avoid the use of repetitive steering points without turning that avoidance 
into an explicit design rule.
\end{enumerate}

\end{document}